\documentclass[aps,nofootinbib,notitlepage,superscriptaddress,10pt,preprint]{revtex4-1}

\usepackage{graphicx}
\usepackage{amsmath,amssymb}
\usepackage{hyperref}
\usepackage{braket}
\usepackage{subfigure}
\usepackage{float}
\usepackage[dvipsnames]{xcolor}
\usepackage{soul}
\usepackage{rotating}
\usepackage{multirow}
\usepackage{mathtools}
\usepackage[normalem]{ulem}
\usepackage{comment}
\usepackage{cprotect}
\usepackage{xspace}

\usepackage{makecell}

\usepackage[margin=0.75in]{geometry}

\hypersetup{
	colorlinks=true,
	linkcolor=blue,
	citecolor=blue,
	urlcolor = black,
}

\usepackage{xcolor}

\def\nn{\nonumber} 
\def\pa{{\partial}}

\def\l{\left}
\def\r{\right}
\def\d{{\rm d}}
\def\th{\tilde{H}}

\def\vx{{\bf x}}
\def\vk{{\bf k}}

\def\th{\tilde{H}}

\def\Mp{M_{\scriptscriptstyle{\rm Pl}}}

\begin{document}
	
\title{Remarks on an extended $R^2$ model}

\author{Amir Ghalee}
\affiliation{{Department of Physics, Tafresh  University,
P. O. Box 39518-79611, Tafresh, Iran}}

\begin{abstract}	
Observation of the cosmic microwave background is consistent with the $R^2$ model. In the $R^2$ model, there exists a dimensionless constant which must be assigned a very large value to reconcile the model with the data. We show that the large value for this parameter can be used to propose an extended model. The scalar spectral index and the tensor-to-scalar ratio of the extended model have been obtained. Also, we obtain the  three-point correlation function of the curvature perturbation to estimate the primordial non-Gaussianities of the extended model. It has been shown that the predictions of the extended model are in agreement with the observations.

\end{abstract}
\maketitle

\section{Introduction}
The latest \textit{Planck} results \cite{{plank1},{plank2},plank3}, provide data to support the cosmic inflation paradigm in which a period of accelerated expansion in the very early Universe has been proposed \cite{{stra},{kaz},{gu},{lin1},{st}}.
The data, which is based on observations of the cosmic microwave background (CMB), have provided constraints on inflationary cosmological models \cite{plank3}.\\
According to the \textit{Plank} results \cite{plank3}, the so-called $R^2$ model( Starobinsky model \cite{stra}) is in good agreement with the observations.\
By setting  $\bar{h}=c=1$, the $R^2$ model is described by  
\begin{equation}\label{i-0}
S=\int d^{4}x\sqrt{-g}\left[\frac{\Mp^2 R}{2}+ f_{s}R^2\right],
\end{equation}
where $\Mp$ is the reduced Plank mass and $f_s$ is a dimensionless constant that by introducing a mass scale, $M$,  is parametrized by $f_s=\frac{\Mp^2}{12M^2}$ \cite{plank3}.\\
Also, the observations imply the following upper bound on the Hubble parameter during inflation, $H_{*}$, \cite{plank3}
\begin{equation}
\frac{H_{*}}{\Mp}<2.5\times 10^{-5}\hspace{.2cm} (\text{95\% C. L.}).
\end{equation}
On the other-hand, the necessary but not sufficient condition to have the cosmic inflation from the action (\ref{i-0}) is that the second term in the action dominates over the first term. This condition with the upper bound on  $H_{*}$ imply that $f_s\geq 5\times 10^{8}$ \cite{{gr2n},{gr3}}.\\
Regarding the observations, many efforts have been made to generalize the $R^2$ model \cite{{gr2},{re3},{re}, {od}}.\\  
In this paper, our aim is to introduce an extended model for the $R^2$ model which has predictions similar to the $R^2$ model. However, the extended model provides a reason for the value of the dimensionless parameter.\\  
In order to achieve this aim, we introduce a scalar field, $\varphi$, and consider the following model 
\begin{equation}\label{i-1}
S=\int d^{4}x\sqrt{-g}\left[\frac{\Mp R}{2}+ f(X) R^2\right], X\equiv\frac{-\partial_\mu \varphi\partial^\mu\varphi}{2M^4}.
\end{equation}
In Ref. \cite{noh1}, the authors consider a model, in a different context, that has similar form as the above action\footnote[1]{It should be mentioned that inspection of Ref. \cite{noh1} shows that although, at first glance, it seems that the authors began with a general action, they did not consider our model.}. However, our aim and discussions are different. Also, we find that the recent observations are so accurate that we need to check some approximations which have been used for the $R^2$ model in Ref. \cite{re} and references therein. \\
Our discussions are based on a scenario which is described in Fig. 1. 
\begin{figure}
 \includegraphics[width=9cm]{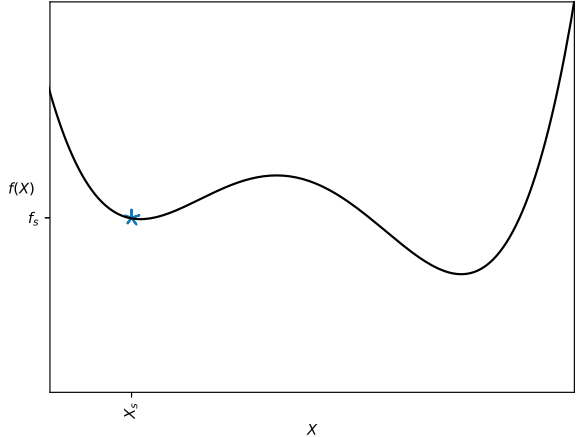}
 \caption{Description of our scenario for the action (\ref{i-1}). $f(X)$ has an arbitrary shape. It has been shown that the model has well-defined behaviour around the $X_s$, which is one of the possible minimums for $f(X)$. It has been shown that one can regard the $R^2$ model as an example of this model for  which we have $f(X_*)\gg1$. We have set $X_s=1$ in the main text.}
\end{figure}
In Fig. 1, $f(X)$ has an arbitrary shape. We want to expand $f(X)$ around a minimum, which is shown by $X_s$. We will show that the model has well-defined behaviour around the $X_s$. Also, we will find that if $f(X_*)\gg1$, the predictions of the model is the same as the $R^2$ model. Moreover, by using data in  Ref. \cite{plank3},  we will obtain a lower bound for $f(X_*)$ which is much lower than $f_s$ in the $R^2$ model.

The organization of this paper is as follows: in Sec. II we study background cosmology of the model. Sec. III is devoted to study the linear perturbation equations and we derive the scalar spectral index and the tensor-to-scalar ratio  of the model. Sec. IV is devoted to study ghost  modes and tachyonic instability.
 In
Sec. V we derive the corresponding  cubic action for scalar metric perturbations to study the primordial non-Gaussianities of the model. Sec. VI is devoted to conclusions.
\section{Background cosmology}
In this section we discus the background cosmology of the action (\ref{i-1}). It is clear that when we consider a constant value for $f(X)$ as $f(X)=f_s$, our results must be reduced to the corresponding results for the $R^2$  model. We will use the excellent review in Ref. \cite{re} to compare our results with the $R^2$ model.\\
As already mentioned, we will study the model close to a minimum of $f(X)$. At first glance, it seems that this approach has limited ranges. However, we will show that, close to the minimum and during inflation, the scalar field is condensed and in Sec. IV we will show that the model is well-defined around the minimum. Then, since $f(X)$ has an arbitrary shape, one can consider it as a function that has many minimums. So, one can choose one of the minimum.   
\subsection{Background equations of motion}
For the background metric, we will use the flat Friedmann-Robertson-Walker( FRW) metric as
\begin{equation}\label{s1_1}
ds^2=-dt^2+a^2\delta_{ij}dx^idx^j,
\end{equation}
where $a=a(t)$ is the scale factor from which the Hubble parameter is defined as
$H=\frac{\dot{a}}{a}$, where dots denote time derivatives.\\
Varying the action (\ref{i-1}) with respect to the metric gives
\begin{align}\label{s1_2}
R_{\mu\nu}-\frac{1}{2}Rg_{\mu\nu}=&2\Mp^{-2}\big[\frac{\pa_\mu\varphi\pa_{\nu}\varphi}{2M^4}f_X R^2
-2RfR_{\mu\nu}\\\nn
&+2\pa_{\mu}R\pa_{\nu}f+2\pa_\nu R\pa_\mu f
+2R\nabla_{\mu}\nabla_{\nu} f\\\nn
&+2f\nabla_\mu\nabla_\nu R
-2f\nabla^2Rg_{\mu\nu}-2R\nabla^2fg_{\mu\nu}\\\nn
&-4\pa^\alpha R\pa_\alpha f g_{\mu\nu}+\frac{fR^2}{2}g_{\mu\nu}\big],
\end{align}
where $f=f(X)$ and $f_X=\frac{df}{dX}$.\\
Using the FRW metric, the time-time  and space-space components of Eq. (\ref{s1_2}) can be obtained as
\begin{equation}\label{s1_3}
3H^2=6H\tilde{H}-\frac{2\Mp^{-2}\l[Xf_X R^2+\frac{fR^2}{2}\r]}{(1+4\Mp^{-2}Rf)},
\end{equation}
and 
\begin{align}\label{s1_4}
-(3H^2+2\dot{H})=&4(\tilde{H}-H)H+4(\tilde{H}-H)^2\\\nn
&+2(\dot{\tilde{H}}-\dot{H})-\frac{\Mp^{-2}fR^2}{(1+4\Mp^{-2}Rf)},
\end{align}
respectively, where
\begin{equation}\label{s1_5}
\tilde{H}=H+\frac{2\Mp^{-2}}{(1+4\Mp^{-2}Rf)}\frac{d}{dt}(Rf).
\end{equation}
From Eqs. (\ref{s1_3}) and (\ref{s1_4}), it turns out that
\begin{equation}\label{s1_6}
\dot{\tilde{H}}=H(\tilde{H}-H)-2(\tilde{H}-H)^2-\frac{\Mp^{-2}Xf_X R^2}{(1+4\Mp^{-2}Rf)}.
\end{equation}
So, from Eq. (\ref{s1_5}) it follows that as far as $1+4\Mp^{-2}Rf>0 $, one can define the following variable 
\begin{equation}\label{s1_7}
\tilde{a}=a(1+4\Mp^{-2}Rf)^{\frac{1}{2}}.
\end{equation}
From the above definition and Eq. (\ref{s1_5}), it follows that  $\tilde{H}=\frac{\dot{\tilde{a}}}{\tilde{a}}$\\
Also, varying the action (\ref{i-1}) with respect to $\varphi$, gives
\begin{equation}\label{s1_8}
\nabla_\mu\l(g^{\mu\nu}\sqrt{-g}f_XR^2\nabla_\nu\varphi\r)=0,
\end{equation}
which leads us to the following equation
\begin{equation}\label{s1_9}
\frac{\dot{X}}{HX}=-6\l(1+\frac{2}{3}\frac{\dot{R}}{HR}\r)\frac{f_X}{f_X+2Xf_{XX}},
\end{equation}
where $f_{XX}=\frac{d^2f}{dX^2}$.
\subsection{Inflationary era }
The slow-roll conditions are characterized  by the
the Hubble flow functions( HFFs), $\epsilon_i$, as
\begin{equation}\label{cond}
 \epsilon_1=-\frac{\dot{H}}{H^2}\ll1,\hspace{.2cm} \epsilon_{n+1}=\frac{\dot{\epsilon}_n}{H\epsilon_n}\ll1,\hspace{.2cm} (n\geq 1).
\end{equation}
According to the \textit{Plank} results, we have \cite{plank3}
\begin{align} 
 &\epsilon_1<0.0063 \hspace{1cm} (0.0039) \hspace{0.3cm} (95\% ~\text{ C. L. }),   \\\nn
  & \epsilon_2=0.030_{-0.005}^{+0.007}\hspace{.2cm} (0.031 \pm 0.005) \hspace{.3cm}     (68\% ~\text{C. L. }).
\end{align}
Also, for any time-dependent quantity such as $X(t)$ we have to impose $\frac{\dot{X}}{HX}\ll1$.\\
It should be mentioned that to report the above data, the Einstein frame  is used \cite{plank3}. In this work, we will use the Jordan frame  \cite{re}. Although we just need to know that during inflation we have $\epsilon_i\ll1$, in a few places we feel that the numerical values for $\epsilon_i$ help us to clarify some issues. In such cases, we will take $\epsilon_1=4\times10^{-3}$.  In other words, we make the strong assumption that during inflation the order of magnitude of $\epsilon_i$ are the same in the both frames. \footnote[2]{The Einstein and Jordan frame are related by a conformal transformation. There is no physical principle that shows invariant of physical quantities under the conformal transformation. For an example, see Eq. (\ref{report_te}).}\\     
Although the slow-roll conditions can be regarded as initial conditions, they must remain valid during inflation. So, the consistency of the conditions in Eq. (\ref{cond}) with the dynamics of the model must be investigated.\\ 
Using $R=12H^2+6\dot{H}$, Eq. (\ref{s1_9}) can be rewritten as
\begin{align}\label{s1_11}
\frac{dX}{d\cal{N}}=\frac{-6Xf_X\l[1-\frac{4}{3}\epsilon_1\l(1+\frac{\epsilon_2}{2(2-\epsilon_1)}\r)\r]}{f_X+2Xf_{XX}},
\end{align}
where ${d\cal{N}} = Hdt=d\ln a$, which measures the number of \textit{e}-folds ${\cal{N}}$ of inflationary expansion. Also,
note that the above equation is in fact exact and in the above equation and all differential equations for $\epsilon_i$ actually we have $\epsilon_i=\epsilon_i(t)$.\\
Now, we want to study the model close to the minimum of $f(X)$ which can be represented as $X=1$. Note that, this expansion is a nontrivial step. We must show that this assumption leads us to a well-defined behaviour for the model. As we will show, our results support this assumption.\\
Although our discussions are general, in what follows one can consider $f(X)=f_s+\frac{{\cal{F}}}{2}(X-1)^2+...$. As an example, if we take the ``Higgs-like`` shape as $f(X)=a_1-a_2 X+a_3X^2$ then $f(X)$ has a minimum at $X_*=\frac{a_2}{2a_3}$ . Then by using $f_s\equiv a_1-\frac{a_2^2}{4a_3}$ and ${\cal{F}}\equiv2a_3$, it turns out that $f(X)=f_s+\frac{{\cal{F}}}{2}(X-X_*)^2$. Note that, at the background level, it is always possible to take $X_*=1$.\\
Expanding Eq. (\ref{s1_11}) to first order in $(X-1)$ gives
\begin{equation}\label{s1_12}
\frac{dX}{d\cal{N}}=-3K(\epsilon_1,\epsilon_2)(X-1),
\end{equation}
where
\begin{equation}\label{s1_13}
K(\epsilon_1,\epsilon_2)\equiv1-\frac{4}{3}\epsilon_1\l(1+\frac{\epsilon_2}{2(2-\epsilon_1)}\r).
\end{equation}
To determine behaviour of $K(\epsilon_1,\epsilon_2)$, expanding Eq. (\ref{s1_3}) to first
order in $X-1$  results in
\begin{align}\label{s1_14}
\frac{1}{24f\Mp^{-2}\epsilon_1 H^2(2-\epsilon_1)}-\frac{3}{2}=&\frac{\dot{\epsilon}_1}{H\epsilon_1(2-\epsilon_1)}\\\nn
&+(X-1)\frac{f_{XX}(2-\epsilon_1)}{f\epsilon_1},
\end{align}
where we have used $R=6H^2(2-\epsilon_1)$ which results in $\dot{R}=-12H^3\epsilon_1(2-\epsilon_1)-6H^2\dot{\epsilon_1}$.\\
Note that, Eq. (\ref{s1_14}) is in fact exact equation around the minimum of $f(X)$ and we just write it in terms of $\epsilon_1$ .\\
From Eq. \ref{s1_14} it turns out that for the $R^2$ model we have to take the following condition to have the slow-roll inflation
\begin{align}\label{s1_af1}
\frac{1}{24f_s\Mp^{-2}\epsilon_1 H^2(2-\epsilon_1)}-\frac{3}{2}\approx \epsilon_1\ll1\\\nn
 (\text {for the $R^2$ model }), 
\end{align}
that can be rewritten as 
\begin{equation}\label{s1_afff}
\frac{1}{4f_s\Mp^{-2}\epsilon_1 R}-\frac{3}{2}\approx\epsilon_1\ll1  \hspace{.3cm}(\text {for the $R^2$ model }).
\end{equation}
To see another parametrization of the above equation in the $R^2$ model see Ref. \cite{rep1}.\\
Since during inflation we have $\epsilon_{i}\ll1$, Eq. (\ref{s1_afff}) results in $\Mp^{-2}Rf\gg1$ but the reverse is not true.
Also, from Ref. \cite{plank3}, we know that $\frac{H}{\Mp}<2.5\times10^{-5}$. Thus, if we take $\epsilon_1\approx 4\times 10^{-3}$ for the $R^2$ model, then from Eq. (\ref{s1_af1}) it follows that $f_s\geq 5\times 10^{8}$.\\
As in the $R^2$ model, we take the following condition as the necessary condition for inflation in our model  
\begin{equation}\label{s1_af}
\frac{1}{24f\Mp^{-2}\epsilon_1 H^2(2-\epsilon_1)}-\frac{3}{2}\equiv g\epsilon_1\ll1, 
\end{equation}
where $g$ is a constant.\\
As noted earlier, at the end of calculations, we have to check such assumptions.\\ 
From Eqs. (\ref{s1_14}) and (\ref{s1_af}) it follows that
\begin{equation}\label{1400}
  \frac{\dot{\epsilon}_1}{H\epsilon_1(2-\epsilon_1)}+(X-1)(2-\epsilon_1)\frac{f_{XX}}{f\epsilon_1}=g\epsilon_1. 
  \end{equation}  
Also, expanding Eq. (\ref{s1_6}) to first order in $(X-1)$  and write the results in terms of $\epsilon_1$ leads us to
\begin{align}\label{s1_150000}
&\frac{\ddot{\epsilon}_1}{2H^2(2-\epsilon_1)}+\frac{\dot{\epsilon}_1}{2H(2-\epsilon_1)}(3-6\epsilon_1)\\\nn
&-\frac{3}{2}\frac{f_{XX}}{f}(X-1)(2-\epsilon_1)-3\epsilon_1^2=\\\nn
&-\frac{\epsilon_1}{24f\Mp^{-2}H^2(2-\epsilon_1)}.
\end{align}
Inserting Eq. (\ref{s1_af}) into Eq. (\ref{s1_150000})  then gives 
\begin{align}\label{s1_15}
&\frac{\ddot{\epsilon}_1}{2H^2(2-\epsilon_1)}+\frac{\dot{\epsilon}_1}{2H(2-\epsilon_1)}(3-6\epsilon_1)\\\nn
&+(g-\frac{3}{2})\epsilon_1^2
-\frac{3}{2}\frac{f_{XX}}{f}(X-1)(2-\epsilon_1)=0,
\end{align}
From Eqs. (\ref{1400}) and (\ref{s1_15}), we can eliminate the terms which are proportional to $(X-1)$ . This procedure yields a differential equation for $\epsilon_1$ as
\begin{equation}\label{s1_16}
\frac{d^2\epsilon_1}{{d\cal{N}}^2}+(6-7\epsilon_1)\frac{d\epsilon_1}{d\cal{N}}-(3+g)\epsilon_1^2(2-\epsilon_1)=0,
\end{equation}
where $d{\cal{N}}=Hdt$ is used.\\
The above equation is an example of the Lienard's equation \cite{perko}. There exists a systematic procedure to study the Lienard's equation. By introducing an auxiliary variable $y$, it is easy to find the following representation for Eq. (\ref{s1_16})
\begin{equation}\label{s1_17}
 \begin{cases} 
   \frac{d\epsilon_1}{d\cal{N}}=-6\epsilon_1+\frac{7}{2}\epsilon_1^2 +y& ,\\
   \frac{dy}{d\cal{N}}=(3+g)\epsilon_1^2(2-\epsilon_1).& 
  \end{cases}
\end{equation}
Eq. (\ref{s1_17}) is an autonomous system and in the phase space of $(\epsilon_1,y)$ has the following fixed points
\begin{equation}\label{s1_18}
(0,0),\hspace*{.2cm}(2,-2).
\end{equation}
During inflation, $\epsilon_1\ll 1$, we have to take initial values around $(0,0)$. So, let us first study this fixed point.\\
To determine behaviour of this fixed point, from Eq. (\ref{s1_17}) we can obtain the linearized system around $(0,0)$ as
\begin{equation}\label{s1_19}
\begin{pmatrix}
\frac{d\epsilon_1}{d\cal{N}}&   \\
\frac{dy}{d\cal{N}} &   
\end{pmatrix}=A_{*}\begin{pmatrix}
\epsilon_1 &   \\
y &   
\end{pmatrix},
\end{equation}
where $A_{*}$ is the stability matrix around $(0,0)$ which has the following form
\begin{equation}\label{s1_20}
A_{*}=\begin{pmatrix}
-6 & 1  \\
0 & 0  
\end{pmatrix}.
\end{equation}
The above matrix has two eigenvalues as $\lambda_1=-6$ and $\lambda_2=0$. So, $(0,0)$ is the stable fixed point.\\
Therefore, if we take $\epsilon_1\ll1$ as the initial condition, this result shows that during inflation $\epsilon_1$ remains small. Note that the stability of $(0,0)$ does not depend on $g$.\\
Furthermore, Eq. (\ref{s1_19}) can be solved and gives the following expression    
\begin{equation}\label{s1_21}
\epsilon_1=C_1e^{-6\cal{N}}+C_2\hspace{.1cm}(\text{ close to $(0,0)$}).
\end{equation}
where $C_1$ and $C_2$ are constants of integration. Therefore
\begin{equation}\label{s1_22}
\epsilon_2=\frac{\dot{\epsilon}_1}{H\epsilon_1}=\frac{1}{\epsilon_1}\frac{d\epsilon_1}{d\cal{N}}=\frac{-6}{1+\frac{C_2}{C_1}e^{6\cal{N}}}\hspace{.1cm}(\text{ close to $(0,0)$}).
\end{equation}
Since ${\cal{N}}=\int H dt$ is the number of $e$-folds of inflationary expansion, one can choose $C_1$ and $C_2$ in such a way that we have $\epsilon_2\ll1$ during inflation. For this goal, it is sufficient to take
\begin{equation}\label{sb1}
C_1,C_2\ll1,\hspace{.2cm}\frac{C_2}{C_1}\gg1.
\end{equation}
Just to complete our knowledge about Eq. (\ref{s1_17}), let us consider the other fixed point in Eq. (\ref{s1_18}), i.e. $(2,-2)$.  The linearized system around $(2,-2)$ is
\begin{equation}
\begin{pmatrix}
\frac{d\epsilon_1}{d\cal{N}}&   \\
\frac{dy}{d\cal{N}} &   
\end{pmatrix}=A_{**}\begin{pmatrix}
\epsilon_1-1 &   \\
y-1 &   
\end{pmatrix},
\end{equation}
where 
\begin{equation}
A_{**}=\begin{pmatrix}
8 & 1  \\
-4(3+g) & 0  
\end{pmatrix}.
\end{equation}
$A_{**}$ has two positive eigenvalues as $\lambda_1=4+2\sqrt{1-g}$  and $\lambda_2=4+2\sqrt{1-g}$. Thus, for any value for $g$, one of the eigenvalues of $A_{**}$ has positive sign. So, $(2,-2)$ is not a stable fixed point.\\
Fig. 2, shows the phase space portrait of Eq. (\ref{s1_17}) for $g=0$, which is in agreement with our discussions.
\begin{figure}[h]
 \includegraphics[width=9cm]{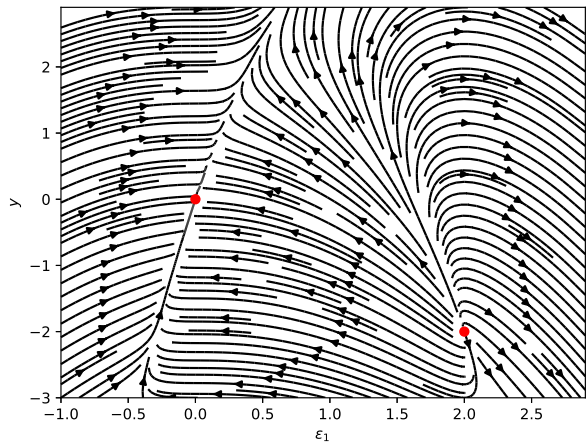}
 \caption{The phase space portrait of Eq. (\ref{s1_17}}) for $g=0$. $(0,0)$ is the stable fixed point and $(2,-2)$ is the unstable fixed point.
\end{figure}
Here, the important result is that if we take initial conditions in such a way that $\epsilon_i\ll1$, the dynamics of the system do not refute such  assumptions.\\
Of course, recall that we have to consider Eq. (\ref{s1_af}) as the essential step that leads us to the stated results. However,  
eventually at some time, say $t=t_f$, the Hubble parameter dropped enough that the assumption made in Eq. (\ref{s1_af}) could not be valid. So, let us check Eq. (\ref{s1_af}).
For this goal, consider $t_i$ as the beginning of inflation 
and $t_f$ as the last time that Eq. (\ref{s1_af}) is valid. Now, we want to determine ${{\cal{N}}_{max}}$ as the maximum  number of \textit{e}-folds for which Eq. (\ref{s1_af}) is valid. \\
To find ${{\cal{N}}_{max}}$,  from  Eq. (\ref{s1_22}) it follows that
\begin{equation}\label{sb2}
\epsilon_1(t_i)\approx C_1+C_2,\hspace{.2cm}\epsilon_1(t_f)\approx C_2.
\end{equation}
On the other hand, since $\epsilon_1=-\frac{\dot{H}}{H^2}=-\frac{1}{H}\frac{dH}{d\cal{N}}$, Eq. (\ref{s1_22}) can be solved in terms of $N$ as
\begin{align}
H(t_f)=&H(t_i)\l(e^{-\frac{C_1}{6}}e^{-C_2{\cal{N}}}\exp[C_2e^{-6\cal{N}}]\r)\\\nn
&\approx H(t_i)e^{-C_2{\cal{N}}}.
\end{align}
Now, note that Eq. (\ref{s1_af}) is valid as far as $H(t_i)\approx H(t_f)$. So, from the above result,  we have to impose the following relation
\begin{equation}\label{sb3}
C_2 {{\cal{N}}_{max}} < \frac{1}{2}.
\end{equation}
Then, from Eqs. (\ref{sb1}), (\ref{sb2}) and (\ref{sb3}) it follows that
\begin{equation}\label{sb4}
{{\cal{N}}_{max}}< \frac{1}{2\epsilon_1(t_f)}\approx\frac{1}{2\epsilon_1(t_i)}.
\end{equation}
Therefore, if we take $\epsilon_1\approx 4\times 10^{-3}$, we have ${{\cal{N}}_{max}}<250$, which is sufficient to solve the horizon problem \cite{weinberg}.\\
Returning now to Eqs. (\ref{s1_12}) and (\ref{s1_13}). Eq. (\ref{s1_12}) can be solved as
\begin{align}\label{s1_23}
X-1=X_{*}\exp{\l(-3\int K(\epsilon_1,\epsilon_2){d\cal{N}}\r)},
\end{align}  
where $X_{*}$  is a constant of integration. From Eqs. (\ref{s1_13}), (\ref{s1_21}), (\ref{s1_22}) and (\ref{s1_23}), it is clear that as time passes then we have $X\rightarrow 1$. So, if as an initial condition we take $X_{*}$ close to $X=1$,
it remains close to $X=1$ during inflation era. Therefore, the scalar filed is condensed during inflation. In the next section we will show that this fact leads us to obtain a nontrivial speed of sound for the scalar perturbations.    
\section{Linear perturbations}
In this section, we study linear perturbations of the model. For this goal a suitable gauge is used. As discus in Ref. \cite{{re},{rador}}, this is a nontrivial task for the modified gravities.
Then,  we expand the action (\ref{i-1}) to second order in scalar and tensor perturbations and study the implications of them for the model.
\subsection{Gauge fixing for the scalar perturbations}
In this part, we will write perturbed equations as
\begin{equation}\label{s2_1}
\delta G_{\mu}^{\nu}=\Mp^{-2}\delta T_{\mu}^{\nu},
\end{equation}
where $G_{\mu\nu}=R_{\mu\nu}-\frac{1}{2}Rg_{\mu\nu}$ is the Einstein tensor, and $\delta T_{\mu\nu}$ is corresponding  perturbed quantities.\\  
To choose the gauge, let us parametrized the scalar part of the perturbed metric as \cite{weinberg}
\begin{align}\label{s2_2}
ds^2=&-\l(1+2\Phi(\vx,t)\r)dt^2+2a\pa_iB(\vx,t)dtdx^i\\\nn
&+\l(1+2\zeta(\vx,t)\r)a^2\delta_{ij}dx^idx^j.
\end{align} 
Also, we define the following variables 
\begin{align}\label{s2_3}
& \tilde{\Phi}(\vx,t)=\Phi(\vx,t)+\frac{2\Mp^{-2}}{1+4\Mp^{-2}Rf}\delta(Rf),\nn\\\
& \tilde{\zeta}(\vx,t)=\zeta(\vx,t)+\frac{2\Mp^{-2}}{1+4\Mp^{-2}Rf}\delta(Rf).
\end{align}
Using Eq. (\ref{s2_2}), the $(0i)$ component of Eq. (\ref{s1_2}) can be obtained as( after drop $\pa_i$ from the both sides)
\begin{align}\label{s2_4}
\tilde{H}\tilde{\Phi}(\vx,t)-\dot{\tilde{\zeta}}=&\frac{2\Mp^{-2}}{1+4\Mp^{-2}Rf}\Big[\frac{Xf_X R^2}{2\dot{\varphi}}\delta\varphi\\\nn
&+3(\tilde{H}-H)\delta(Rf)\Big].
\end{align}
Also, by using Eq. (\ref{s2_2}), the $(ii)$ component of Eq. (\ref{s1_2}) yields 
\begin{align}\label{s2_5}
&(3\th^2+2\dot{\th})\tilde{\Phi}(\vx,t)+\th\dot{\tilde{\Phi}}(\vx,t)-\ddot{\tilde{\zeta}}(\vx,t)\\\nn
&-3\th\dot{\tilde{\zeta}}(\vx,t)=\Big[\Mp^{2}(4H\th-3H^2-\th^2)\tilde{\Phi}(\vx,t)\\\nn
&+\big(8H\th+4\th^2-3\dot{H}-8H^2+8\dot{\th}\big)\delta(Rf)\\\nn
&-\frac{Rf\delta(R)}{2}+6(\th-H)\frac{d\delta(Rf)}{dt}\Big]\frac{\Mp^{-2}}{1+4\Mp^{-2}Rf}\\\nn
&-(\th-H)\dot{\tilde{\zeta}}(\vx,t),
\end{align}
where Eq. (\ref{s1_6}) is used to obtain the above result.\\
As discuss in Ref. \cite{weinberg}, one can define the scalar component of velocity perturbation, $\delta u$, as $\delta G^{0}_{i}=\Mp^{-2}(\bar{\rho}+\bar{p})\pa_i \delta u$. Then the comoving gauge is define as the gauge in which $\delta u=0$. So, from Eq. (\ref{s2_4}), the comoving gauge can be imposed by the following condition
 \begin{equation}\label{s2_6}
\frac{Xf_X R^2}{2\dot{\varphi}}\delta\varphi+3(\tilde{H}-H)\delta(Rf)=0.
   \end{equation}  
Therefore, in the comoving gauge, we have $\dot{\tilde{\zeta}}(\vx,t)=\th\tilde{\Phi}(\vx,t)$. Substituting this into left-hand side( L. H. S) of Eq. (\ref{s2_5}) results in 
   \begin{equation}\label{s2_7}
   \text{L. H. S of Eq. (\ref{s2_5})}= \dot{\tilde{H}}\tilde{\Phi}(\vx,t).
   \end{equation}
To obtain the right-hand side( R. H. S) of Eq. (\ref{s2_5}), note that    
\begin{equation}\label{s2_8}
\delta(Rf)=Rf_X+f\delta^{1}R,\hspace{.1cm}\delta X=-2X\Phi(\vx,t)+2X\frac{\delta\dot{\varphi}}{\dot{\varphi}},
\end{equation}
where $\delta^1R$ is linear part of the perturbed $R$.
So, 
\begin{equation}\label{s2_9}
fR\delta^1R=R\delta(Rf)+2R^2Xf_X\l(\tilde{\Phi}(\vx,t)-\frac{\delta\dot{\varphi}}{\dot{\varphi}}\r).
\end{equation}
Then, by taking the time derivative of Eq. (\ref{s2_6}) and using Eq. (\ref{s1_9}) , we have
\begin{align}\label{s2_10}
&-\frac{Xf_XR^2}{6}\frac{\delta\dot{\varphi}}{\dot{\varphi}}=(\dot{\th}-\dot{H})\delta(Rf)\\\nn
&+(\th-H)\frac{d}{dt}\delta(Rf)
+3H(\th-H)\delta(Rf).
\end{align}
By eliminating $\delta\dot{\varphi}$ from  Eq. (\ref{s2_9}) and Eq. (\ref{s2_10}) we have
\begin{align}\label{s2_11}
fR\delta^1R=&\big[R+36H(\th-H)+12(\dot{\th}-\dot{H})\big]\delta(Rf)\\\nn
&+2R^2Xf_X\Phi(\vx,t)+12(\th-H)\frac{d}{dt}\delta(Rf).
\end{align}
Now, using  $\dot{\tilde{\zeta}}(\vx,t)=\th\tilde{\Phi}(\vx,t)$ and by inserting the above result into R. H. S of Eq. (\ref{s2_5}), we have
\begin{align}\label{s2_12}
 \text{R. H. S of Eq. (\ref{s2_5})}=&\dot{\th}\tilde{\Phi}(\vx,t)\\\nn
 &+2\Mp^{-2}(H^2+\dot{H})\delta(Rf).
\end{align}   
Comparing the above result with Eqs. (\ref{s2_7}) and (\ref{s2_6}) gives the following results
\begin{equation}\label{s2_13}
\delta(Rf)=0,\hspace{.2cm} \delta\varphi=0.
\end{equation}
So, one can use Eq. (\ref{s2_13}) as the comoving gauge in our model.\\
Regarding the definitions of $\tilde{\zeta}(\vx,t)$, $\tilde{\Phi}(\vx,t)$  
 and Eq. (\ref{s2_13}), we will use $\zeta(\vx,t)$ and $\Phi(\vx,t)$ in the following sections. 
\subsection{Linear scalar perturbations from the second order action}
In the comoving gauge, $\zeta(\vx,t)$ is the gauge invariant quantity \cite{weinberg}. So it is convenient to obtain a differential equation in terms of $\zeta(\vx,t)$. Also, note that we have shown that it is possible to use a gauge in which $\delta\varphi=0$. So, as the background level, this property help us to set $X_*=1$.\\
To obtain this equation, one way is to use the perturbed $(00)$ component of Eq. (\ref{s1_2}). Using Eqs. (\ref{s1_3}) (\ref{s1_6}) and (\ref{s2_12}) it turns out that the perturbed $(00)$ component of Eq. (\ref{s1_2}) is 
\begin{align}\label{s2_14}
&\tilde{H}a\frac{\pa^2B(\vx,t)}{a^2}+\frac{\pa^2\zeta(\vx,t)}{a^2}=\\\nn
&\frac{a^2(1+4\Mp^{-2}Rf)}{\th^2}\Sigma\dot{\zeta}(\vx,t),
\end{align}
where
\begin{align}\label{s2_15}
\Sigma\equiv & 3(\th-H)^2+\frac{2X\Mp^{-2}R^2}{1+4\Mp^{-2}Rf}\Big[\frac{f_X}{2}+Xf_{XX}\\\nn
&-\frac{2Xf_X^2}{f}\Big].
\end{align}
From Eq. (\ref{s2_14}) and $\dot{\zeta}(\vx,t)=\th\Phi(\vx,t)$, it is easy task to obtain an equation  for $\zeta(\vx,t)$, as is shown by Eq. (\ref{s2_27}).\\
However, we need the second order action in the next section. Thus, we shall obtain the differential equation for $\zeta(\vx,t)$ by using the second order action.\\
For this goal, the 
Arnowitt-Deser-Misner (ADM) formalism is used~\cite{arnowitt-1960}. In the ADM formalism, to foliate the spacetime with spacelike hypersurfaces, a unite normal vector, $n^\alpha$, is used and
the metric is parametrized as~\cite{arnowitt-1960}
\begin{equation}\label{s2_16}
\d s^2 = -N^2dt^2 
+ h_{ij}\, \l(N^{i}dt
+ \d x^{i}\r)\, \l(N^{j}dt +\d x^{j}\r),
\end{equation}
where~$N$, $N^{i}$, $h_{ij}$ are the lapse, shift, spatial metric respectably. Also, in the ADM formalism we have the following relations~\cite{{arnowitt-1960},{wald}}
\begin{align}\label{s2_17}
&\sqrt{-g}=N\sqrt{h},\\\nn
& R=^{(3)}R+K^{ij}K_{ij}-K^2-2\big(n^{\alpha}_{;\beta}n^{\beta}-n^{\alpha}n^{\beta}_{;\beta}\big)_{;\alpha},
\end{align}
where $;$ denotes the covariant derivative and $^{(3)}R$
is the Ricci scalar constructed from $h_{ij}$ and ~\cite{{arnowitt-1960},{wald}}
\begin{equation} \label{s2_18}
K_{ij} =
\frac{1}{2N}\,\l[\dot{h}_{ij}
- \l({}^{(3)}\nabla_i\, N_j + {}^{(3)}\nabla_j\, N_i\r)\r], 
\end{equation}
while $K = h_{ij}K^{ij}$ and ${}^{(3)}\nabla$ denotes covariant derivative with respect to $h_{ij}$.\\
Following \cite{mald}, we will use $ h_{ij}=a^2e^{2\zeta(\vx,t)}\delta_{ij}$ and expand the shift and laps as
\begin{align}\label{s2_19}
& N=1+\alpha_1+\alpha_2+...,\hspace{.2cm}N_i=\pa_i\psi +\beta_i,\\\nn
&\psi=\psi_1+\psi_2+...,\hspace{.2cm}\beta_i=\beta_i^{(1)}+\beta_i^{(1)}+...\hspace{.2cm},
\end{align}
where $\pa_i\beta_i=0$.
Also, we will use the following expansion for the Ricci scalar
\begin{equation}\label{s2_20}
R=\bar{R}+\delta R=\bar{R}+\delta^1R+\delta^2R+...\hspace{.2cm},
 \end{equation} 
where $\bar{R}$ is the background value of the Ricci scalar and  $i$ in $\delta^iR$ denotes the order of perturbations. The explicit form of the above quantities are given by Eqs. (\ref{apbb1}) and (\ref{apbb2}) in appendix A.\\
From Eqs. (\ref{s2_2}), (\ref{s2_16}) and Eq. (\ref{s2_19}), it follows that
\begin{equation}\label{s2_21}
\Phi(\vx,t)=\alpha_1,\hspace{.2cm}aB(\vx,t)=\psi_1.
\end{equation}
So, the action (\ref{i-1}) becomes
\begin{align}\label{s2_22}
S= \frac{\Mp^2}{2}\int &\Big[a^3e^{3\zeta(\vx,t))}N\Big]\Big[^{(3)}R+K^{ij}K_{ij}-K^2\\\nn
&+2\Mp^{-2}f(X)\l(\bar{R}+\delta^1R+\delta^2R\r)^2\Big].
\end{align}
Now, using the comovig gauge $\dot{\zeta}(\vx,t)=\th\Phi(\vx,t)$, and Eq. (\ref{s2_13}) and using the fact that
\begin{equation}\label{s2_23}
\frac{1}{N^2}=1-2\frac{\dot{\zeta}(\vx,t)}{\th}-2\alpha_2+3\frac{\dot{\zeta}^2(\vx,t)}{\th^2}+...\hspace{.2cm},
\end{equation}
we have
\begin{align}\label{s2_24}
f(X)=&f(\bar{X})+\bar{X}\Big(-2\frac{\dot{\zeta}(\vx,t)}{\th}-2\alpha_2\\\nn
&+3\frac{\dot{\zeta}^2(\vx,t)}{\th^2}\Big)+2\bar{X}^2\frac{\dot{\zeta}^2(\vx,t)}{\th^2}+...\hspace{.2cm},
\end{align}
where $\bar{X}=\frac{\dot{\varphi}^2}{2M^4}$.\\
Plugging the above result into the action (\ref{s2_22}) and using Eqs. (\ref{s2_13}), (\ref{apbb1}), (\ref{apbb2}) and $\dot{\zeta}(\vx,t)=\th\Phi(\vx,t)$ , leads us to the second order action, $\delta^2S$, as 
\begin{align}\label{s2_25}
\delta^2S=\int & dt d^3x a^3\Mp^2\Big[\frac{a^2\Sigma}{\th^2}\dot{\zeta}^2(\vx,t)\\\nn
&+(1+4\Mp^{-2}Rf)(\frac{\dot{\th}}{\th^2}+\frac{H}{\th}-1)\frac{(\pa\zeta(\vx,t))^2}{a^2}\Big],
\end{align}
where we have used integration by parts and dropping surface terms.\\
Note that, one can use Eq. (\ref{s1_6}) to find that
\begin{equation}\label{s2_26}
 \th-H-\frac{\dot{\th}}{\th}=\frac{1}{\th}\l(3(\th-H)^2+\frac{\Mp^{-2} Xf_XR^2}{1+\Mp^{-2}Rf}\r).
  \end{equation} 
Variation of Eq. (\ref{s2_25}) with respect to $\zeta(\vx,t)$, and using Eq. (\ref{s2_26}), results in 
\begin{equation}\label{s2_27}
\ddot{\zeta}_k+\frac{\dot{w}_s}{w_s}\dot{\zeta}_k+\frac{k^2}{a^2}c_s^2\zeta_k=0,
\end{equation}
where
\begin{equation}\label{s2_28}
\zeta_k=\int d^3\vx\zeta(\vx,t)e^{-i\vx.\vk},\hspace{.2cm}w_s\equiv a^3\frac{1+4\Mp^{-2}Rf}{\th^2}\Sigma,
\end{equation}
and $c_s$ is the speed of sound, which has the following form
\begin{align}\label{s2_29}
&c_s^2\equiv\frac{1}{1+\Xi},\\\nn &\Xi\equiv\frac{2M_p^{-2}X^2R^2\l(f_{XX}-2f_X^2\r)}{3(1+4\Mp^{-2}Rf)(\th-H)^2+\Mp^{-2}Xf_XR^2}.
\end{align}
From Eq. (\ref{s2_29}), it is clear that for the $R^2$ model, we have $c_s=1$. Note that at the minimum, $f_{X}=0$, $f_{XX}>0$, the above equation yields $\Xi>0$ and $0<c_s^2\leq1$. Also, Eq. (\ref{s2_29}) shows that higher order terms in the expansion  around the minimum of $f(X)$, are suppress not only by powers of $(X-1)$ but also by $\frac{H}{\Mp}$. Since in Eq. (\ref{s2_29}) everything is exact, this property shows that the model is well-defined around the minimum of $f(X)$.\\
Now, let us focus on the inflationary era. Expanding  Eq. (\ref{s2_29}) around the minimum of $f(X)$, i.e. $X=1$, and keep the terms proportional to $\epsilon_1$ and $\epsilon_2$ results in
\begin{equation}\label{s2_30}
\Xi=\frac{f_{XX}}{f\epsilon_1^2}\l(2+2\epsilon_1-\epsilon_2\r)+...,
\end{equation}
where Eq. (\ref{s1_af}) is used.\\
For reasons that will soon become clear, we have to impose the following condition
\begin{equation}
\frac{f_{XX}}{f\epsilon_1^2}\ll1.
\end{equation}
Thus, $\Xi\ll1$. So, during inflation, we have 
\begin{equation}\label{s2_31}
c_s^2=1-\frac{f_{XX}}{f\epsilon_1^2}\l(2+2\epsilon_1-\epsilon_2\r).
\end{equation}
Furthermore, it is convenient to define the following variable
\begin{equation}
s=\frac{\dot{c}_s}{Hc_s}.
\end{equation}
So, from Eq. (\ref{s2_31}), it follows that
 \begin{equation}
s\approx2\epsilon_2\frac{f_{XX}}{f\epsilon_1^2}\ll1.
\end{equation}
For fluctuations outside the horizon, $\frac{k}{aH}\ll1$, Eq. (\ref{s2_27}) becomes
\begin{equation}\label{s2_32}
\ddot{\zeta}_k+\frac{\dot{w}_s}{w_s}\dot{\zeta}_k\approx0,
\end{equation}
which has two solutions as
\begin{equation}\label{s_33}
\zeta_k=C_3,\hspace{.2cm}\zeta_k=C_4\int\frac{dt}{w_s},
\end{equation}
where $C_3$ and $C_4$ are constants of integration.\\
From Eq. (\ref{s2_15}) it is clear that around the minimum of $f(X)$, we have $\Sigma>0$. therefore, if we take  the following condition
\begin{equation}
 1+4\Mp^{-2}Rf>0,
 \end{equation} 
then from Eq. (\ref{s2_28}), it follows that $w_s>0$. Note that, the above condition is consistence with Eq. (\ref{s1_af}).  
Therefore, from Eq. (\ref{s_33}) it turns out that one of the solutions of Eq. (\ref{s2_32}) is constant and the other decays.\\ 
To obtain the power spectrum of $\zeta_k$, it is convenient to represent Eq. (\ref{s2_25}) in the canonical form. For this goal, we define the following variable
\begin{equation}\label{s2_34}
z^2\equiv\frac{2a^2\l(1+4\Mp^{-2}Rf\r)}{\th^2}\Sigma,
\end{equation}
and the conformal time as $d\tau=\frac{dt}{a}$. Then Eq. (\ref{s2_25}) takes the following form
\begin{align}\label{s2_35}
\delta^2S=\Mp^2 &\int d^3xd\tau\Big[\frac{z^2}{2}(\zeta^\prime(\vx,\tau))^2\\\nn
&+a^2(1+4\Mp^{-2}Rf)(\frac{\dot{\th}}{\th^2}+\frac{H}{\th}-1)(\pa\zeta(\vx,\tau))^2\Big],
\end{align}
where a prime $^\prime$  denotes a derivative with respect to $\tau$. Then variation of Eq. (\ref{s2_35}) with respect to $\zeta(\vx,\tau)$ and using the corresponding Fourier component $\zeta_k$, yields
\begin{equation}\label{s2_36}
\zeta^{\prime\prime}_k+2\frac{z^\prime}{z}\zeta^{\prime}_k+c_s^2k^2\zeta_k=0.
\end{equation}
Using Eqs. (\ref{s2_15}), (\ref{s2_26}), (\ref{s2_34}), and imposing Eq. (\ref{s1_af}), gives
\begin{equation}\label{s2_37}
\zeta^{\prime\prime}_k+2aH\l(1+\frac{\epsilon_2}{2}-s\r)
 \zeta^{\prime}_k+c_s^2k^2\zeta_k=0,
\end{equation} 
where we keep the terms proportional to $\epsilon_i$.\\
Now, we want to use the so-called slow-roll approximation which asserts that during inflation era, we have \cite{weinberg}
\begin{equation}\label{s2_38}
aH\approx\frac{-1}{(1-\epsilon_1)\tau}.
\end{equation}
Plug Eq. (\ref{s2_38}) into Eq. (\ref{s2_37}), and keeping only terms of first order in perturbations, results in
\begin{equation}\label{s2_39}
\zeta^{\prime\prime}_k-\frac{2}{\tau}\l(1+\epsilon_1-s+\frac{\epsilon_2}{2}\r)
 \zeta^{\prime}_k+c_s^2k^2\zeta_k=0.
\end{equation}
To solve the above equation, consider a new variable $v_k\equiv \Mp z\zeta_k$ and then using Eq. (\ref{s2_34}) to obtain the following equation
\begin{equation}\label{s2_40}
v^{\prime\prime}_k+\l(c_s^2k^2-\frac{\nu_s^2-\frac{1}{4}}{\tau^2}\r)v_k=0,
\end{equation}
where
\begin{equation}\label{s2_41}
\nu_s\equiv\frac{3}{2}+\epsilon_1+\frac{\epsilon_2}{2}-s.
\end{equation}
Eq. (\ref{s2_40}) has two solutions as $\sqrt{-\tau}H_{\nu_s}^{(1)}(-c_sk\tau)$ and
$\sqrt{-\tau}H_{\nu_s}^{(2)}(-c_sk\tau)$, where $H_{\nu_s}^{(1)}$ and $H_{\nu_s}^{(2)}$ are the Hankel functions for which we have $H_{\nu_s}^{(1)}=H_{\nu_s}^{(2)*}$.\\
To find suitable combinations of the solutions, note that the asymptotic expansion of the Hankel function has the following form \cite{hankel}
\begin{align}\label{s2_42}
&\lim_{k\tau\rightarrow-\infty}H_{\nu_s}^{(1)}(-c_sk\tau)\\\nn
&=
\sqrt{\frac{2}{\pi}}\frac{1}{\sqrt{-c_sk\tau}}e^{-ic_sk\tau}e^{-i\frac{\pi}{2}(\nu_s+\frac{1}{2})}.
\end{align}
If we take the Bunch-Davies vacuum for the perturbations deep inside the horizon, i.e. ${k\tau\rightarrow-\infty}$, the suitable solution is
\begin{equation}\label{s2_43}
v_k=\sqrt{\frac{\pi}{2}}e^{i\frac{\pi}{2}(\nu_s+\frac{1}{2})}\sqrt{-\tau}H_{\nu_s}^{(1)}(-c_sk\tau).
\end{equation}
For outside the horizon limit, i.e. ${k\tau\rightarrow 0}$, 
from Eq. (\ref{s2_43}) it follows that \cite{hankel}
\begin{equation}\label{s2_44}
v_k\approx\frac{i}{\pi}\Gamma(\nu_s)\sqrt{\frac{\pi}{2}}e^{i\frac{\pi}{2}(\nu_s+\frac{1}{2})}\sqrt{-\tau}\l(\frac{-c_sk\tau}{2}\r)^{-\nu_s}.
\end{equation}
The power spectrum of $v_k$ is defined as
\begin{equation}\label{s2_45}
<v_\vk v_\vk'>={P_{v}}\delta(\vk+\vk').
\end{equation}
Also, from $v_k=\Mp z\zeta_k$, it follows that the power spectrum of curvature perturbations, $P_{\zeta}$, can be obtained by  $P_{\zeta}=\Mp^{-2}z^{-2}P_{v}$. Also, the dimensionless scalar power spectrum, $\Delta^2_s$, is defined as 
\begin{equation}
\Delta^2_s=\frac{k^3 P_{\zeta}}{2\pi^2}.
\end{equation}
So, from Eqs. (\ref{s2_34}), (\ref{s2_44}) and (\ref{s2_45}), we have
\begin{equation}\label{s2_46}
\Delta^2_s=\frac{\Mp^{-2}}{4\pi^2}k^3\Gamma^2(\nu_s)z^{-2}(-\tau)\l(\frac{-c_sk\tau}{2}\r)^{-2\nu_s}.
\end{equation} 
Therefor, The scalar spectral index, $n_s$, can be obtained as
\begin{equation}\label{s2_47}
n_s-1=\frac{d\ln(\Delta^2_s)}{d\ln k}=3-2\nu_s,
\end{equation}
which gives
\begin{equation}\label{s2_48}
n_s-1=-2\epsilon_1-\epsilon_2+2s=-2\epsilon_1-\l(1-4\frac{f_{XX}}{f\epsilon_1^2}\r)\epsilon_2.
\end{equation}
If we take $f_{XX}=0$, the above result must be reduced to the corresponding result for the $R^2$ model.\\
The \textit{Plank} results are in good agreement with the $R^2$ model. So, we have to impose $4\frac{f_{XX}}{f\epsilon_1^2}<10^{-2}$. Using $\epsilon_1\approx 4\times10^{-3}$, it turns out that
\begin{equation}
\frac{f_{XX}}{f}<4\times10^{-8}.
\end{equation}
Note that this condition is satisfied if we take $f(X=1)\geq 5\times10^{8}$, and $f_{XX}\approx {{\cal{O}}(1)}$.\\      
\subsection{Tensor perturbations from the second order action}
For the tensor perturbation, we take the following metric \cite{weinberg}
\begin{equation}\label{s2_49}
ds^2=-dt^2+a^2(\delta_{ij}+h_{ij}(\vx,t))dx^idx^j,
\end{equation}
where $h_i^i=\pa_ih_{ij}=0$.\\
The tensor perturbations have two polarization modes, $+,\times$ and the Fourier representations as \cite{weinberg}
\begin{equation}\label{s2_50}
 h_{ij}(\vx,t)=\int \frac{d^3\vk}{(2\pi)^{\frac{3}{2}}}\sum_{\alpha=+,\times}\eta_{ij}^{\alpha}h_{\vk,\alpha}(t)e^{i\vk.\vx},
 \end{equation}
where $\eta^{\alpha}_{ij}\eta^{\alpha'}_{ij}=2\delta_{\alpha\alpha'}$ and $\eta^{\alpha}_{ii}=k^{i}\eta^{\alpha}_{ij}=0$.\\
Plug Eq. (\ref{s2_49}) into the action (\ref{i-1}) and using Eqs. (\ref{s1_3}) and (\ref{s1_6}), results in
\begin{align}\label{s2_51}
\delta^2S_T=\frac{\Mp^2}{8}\int& dtd^3x a^3(1+4\Mp^{-2}Rf)\Big[\\\nn
&\dot{h}^2_{ij}(\vx,t)
-\frac{(\pa h_{ij}(\vx,t))^2}{a^2}\Big],
\end{align}
where we have used integration by parts and dropping surface terms and the subscript $T$ stands for ''Tensor''.\\
Variation of Eq. (\ref{s2_51}) with respect to $h_{ij}(\vx,t)$ and then using Eq. (\ref{s2_50}), yields
\begin{equation}\label{s2_52}
\ddot{h}_{\vk,\alpha}+\frac{\dot{w_T}}{w_T}\dot{h}_{\vk,\alpha}+k^2h_{\vk,\alpha}=0,
\end{equation}
where
\begin{equation}
w_T\equiv 1+4\Mp^{-2}Rf.
\end{equation}
From Eq. (\ref{s2_52}) it is clear that for the fluctuations outside the horizon, $\frac{k}{aH}\ll1$, we have
\begin{equation}\label{s2_53}
\ddot{h}_{\vk,\alpha}+\frac{\dot{w_T}}{w_T}\dot{h}_{\vk,\alpha}\approx0,
\end{equation}
which has two solutions as
\begin{equation}\label{s2_54}
h_{\vk,\alpha}=C_5,\hspace{.2cm}h_{\vk,\alpha}=C_6\int\frac{dt}{w_T},
\end{equation}
where $C_5$ and $C_6$ are constants of integration. Since $w_T>0$, the above result shows that one solution is constant while the other decays.\\ 
To obtain the power spectrum, we define the following variable
\begin{equation}\label{s2_55}
z_T^2\equiv\frac{a^2}{4}(1+4\Mp^{-2}Rf),
\end{equation}
then using the conformal time to obtain the following form for Eq. (\ref{s2_51})
\begin{equation}\label{s2_56}
\delta^2S_T=\frac{\Mp^2}{2}\int d\tau d^3x z_T^2(\l[(h'_{ij}(\vx,\tau))^2-(\pa h_{ij}(\vx,\tau))^2\r].
\end{equation}
Variation of Eq. (\ref{s2_56}) with respect to $h_{ij}(\vx,\tau)$ 
and using the corresponding Fourier component, $h_{k\alpha}$, results in
\begin{equation}\label{s2_57}
 h_{k,\alpha}^{\prime\prime}+2\frac{z^\prime_T}{z_T}{h}_{k,\alpha}^{\prime}+k^2{h}_{k,\alpha}=0.
 \end{equation} 
Using  Eq. (\ref{s2_55}), and $v_{k\alpha}\equiv\Mp z_Th_{k\alpha}$, yields
\begin{equation}\label{s2_58}
v^{\prime\prime}_{k\alpha}+\l[k^2-2a^2H^2(1-2\epsilon_1)\r]v_{k\alpha}=0,
\end{equation}
where Eq. (\ref{s1_af}) is used and only terms of first order in perturbations have been kept.\\
Using Eqs. (\ref{s2_38}) and (\ref{s2_58}) then give, to first order in $\epsilon_i$,
\begin{equation}\label{s2_59}
v^{\prime\prime}_{k\alpha}+(k^2-\frac{2}{\tau^2})v_{k\alpha}=0.
\end{equation}
The above equation has two solutions as $\sqrt{-\tau}H_{\frac{3}{2}}^{(1)}(-k\tau)$, and $\sqrt{-\tau}H_{\frac{3}{2}}^{(2)}(-k\tau)$.\\
From Eq. (\ref{s2_42})  and using the Bunch-Davies vacuum, it turns out that the the suitable solution is
\begin{equation}\label{s2_60}
v_{k\alpha}=\sqrt{\frac{\pi}{2}}e^{-i\pi}\sqrt{-\tau}H_{\frac{3}{2}}^{(1)}(-k\tau).
\end{equation}
Thus, for the outside the horizon limit, i.e. $k\tau\rightarrow0$, the above result gives
\begin{equation}\label{s2_61}
v_{k\alpha}\approx\frac{i}{\pi}\Gamma(\frac{3}{2})\sqrt{\frac{\pi}{2}}e^{-i\pi}\sqrt{-\tau}\l(\frac{-k\tau}{2}\r)^{-\frac{3}{2}}.
\end{equation}
The power spectrum of $v_{k\alpha}$, $P_{v_{k\alpha}}$, is defined by
\begin{equation}\label{s2_62}
<v_{\vk\alpha}v_{\vk\alpha}>={P_{v_{k\alpha}}}\delta(\vk+\vk^{\prime}).
\end{equation}
Since we have defined, $v_{\vk\alpha}=\Mp z_T h_{\vk\alpha}$, it follows that the power spectrum of $h_{\vk\alpha}$ can be obtained by $P_{h_{k\alpha}}=\Mp^{-2}z_T^{-2}P_{v_{k\alpha}}$.
Also, the dimensionless tensor power spectrum, $\Delta^2_T$, is defined as 
$\Delta^2_T=2\frac{k^3 P_{h_{k\alpha}}}{2\pi^2}$.
So,
\begin{equation}\label{s2_63}
\Delta^2_T=k^3\frac{\Mp^{-2}}{\pi^2}z_T^{-2}P_{v_{k\alpha}}.
\end{equation}
 From Eqs. (\ref{s2_61}), (\ref{s2_61}) and (\ref{s2_63}) it follows that
 \begin{equation}\label{s2_64}
 \Delta^2_T=\frac{\Mp^{-2}}{2\pi^2}\Gamma^2({\frac{3}{2}})k^3{z_T^{-2}}(-\tau)(\frac{-k\tau}{2})^{-3}.
\end{equation} 
Therefore, the spectral index for the tensor perturbations, $n_T$, is given by 
\begin{equation}\label{s2_65}
n_T=\frac{d\ln (\Delta^2_T)}{d\ln k},
\end{equation}
which gives
\begin{equation}\label{report_te}
n_T=0  \hspace{0.3cm} (\text{in the leading order for $\Delta^2_T$)}.
\end{equation}
Note that, we use the Jordan frame. As is shown in  Ref. \cite{re}, the corresponding results for the $R^2$ model gives the same results in the Jordan frame \footnote[3]{For the $R^2$ model, it is easy to define the Einstein frame and then obtain $n_T$ in it. To see this point, in this footnote, consider $f=f_s$.       
From Eqs. (\ref{s1_7}), (\ref{s2_55}) it turns out that $z_T=\frac{\tilde{a}}{2}$ and then Eq. (\ref{s2_57}) takes the following form
\begin{equation*}
 h_{k,\alpha}^{\prime\prime}+2\tilde{H}{h}_{k,\alpha}^{\prime}+k^2{h}_{k,\alpha}=0.
 \end{equation*}   
 This equation is the same as the equation for the tensor perturbations of a usual scalar filed in the Einstein frame.}. However, note that the last result is based on the fact that $\Delta^2_T$ dose not  depend on $k$
in the leading order. As is shown in  Ref. \cite{re2}, if one wants to check consistency relation in a modified gravity, sometimes it is necessary to  go beyond the leading order. As we will see, to obtain the tensor-to-scalar ratio ( which is an important observable quantity), we just need to consider our results in the leading order
\subsection{The tensor-to-scalar ratio}
The tensor-to-scalar ratio, $r$, is defined  by
\begin{equation}\label{s2_66}
r=\frac{\Delta^2_T}{\Delta^2_s}|_{k=Ha},
\end{equation}
where $k=Ha$ shows that we evaluate the tensor-to-scalar ratio
at the moment of horizon crossing.\\ 
Using Eqs. (\ref{s2_34}), (\ref{s2_38}) and (\ref{s2_46}), it follows that
\begin{equation}\label{s2_67}
 \Delta^2_s|_{k=Ha}=\frac{1}{12\pi^2(1+4\Mp^{-2}Rf)}\frac{c_s^{-1}}{\epsilon_1^2}\frac{H^2}{\Mp^2},
 \end{equation} 
where we have set $\nu_s\approx \frac{3}{2}$ and the fact that $\Gamma(\frac{3}{2})=\frac{\sqrt{\pi}}{2}$.\\
Using Eq. (\ref{s1_af}), the above result gives
\begin{equation}\label{s2_68}
 \Delta^2_s|_{k=Ha}=\frac{1}{288\pi^2f}\frac{c_s^{-1}}{\epsilon_1^2}.
 \end{equation} 
Also, Eq. (\ref{s2_64}), gives
\begin{equation}\label{s2_69}
 \Delta^2_T|_{k=Ha}=\frac{4}{\pi^2(1+4\Mp^{-2}Rf)}\frac{H^2}{\Mp^2},
 \end{equation}
 which leads us to the following result 
 \begin{equation}\label{s2_70}
 \Delta^2_T|_{k=Ha}=\frac{1}{6\pi^2f}.
 \end{equation}
Finally, Eqs. (\ref{s2_66}), (\ref{s2_68}) and (\ref{s2_70}), results in
\begin{equation}\label{s2_71}
 r=48\epsilon_1^2c_s=48\epsilon_1^2\l(1-\frac{f_{XX}}{f\epsilon_1^2}\r),
 \end{equation} 
where Eq. (\ref{s2_31}) is used.
\section{Ghost  modes and tachyonic instability }
Since the model has extra degrees of freedom, 
in this section we will obtain conditions for which such extra degrees of freedom are not ghost and tachyons. We first focus on the scalar perturbations.\\
To avoid ghost modes in the model, the second order action in Eq. (\ref{s2_25}) gives the following conditions
\begin{equation}\label{report_1}
\Sigma>0,\hspace{.2cm}-(1+4\Mp^{-2}Rf)(\frac{\dot{\th}}{\th^2}+\frac{H}{\th}-1)>0.
 \end{equation}
Note that, Eq. (\ref{report_1}) gives the general conditions to have well-defined model for all stages of cosmological evolution.\\
Although in any stage of cosmological evolution one can use Eqs. (\ref{s2_15}), (\ref{s2_26}) and Eq. (\ref{report_1}) to impose conditions on the parameters of the model, maybe it is better to work in a general framework. For this goal, Eqs. (\ref{s2_15}), (\ref{s2_26})  and Eq. (\ref{report_1}) suggest the following conditions
\begin{equation}\label{report_2}
\frac{f_X}{2}+Xf_{XX}-\frac{2Xf_X^2}{f}\geq 0,1+4\Mp^{-2}Rf>0, f_X\geq0. 
\end{equation}
It is interesting that close to the minimum of $f$, that for which we can use $f=f_s+\frac{{\cal{F}}}{2}(X-1)^2+...$, the above conditions are satisfied if $f_s\gg1$. Note that this condition is consistence with our motivations and our goals in this paper.\\ 
To avoid tachyonic instability of the perturbations, Eq. (\ref{s2_29}) gives the following conditions to have well-defined behavior
\begin{equation}
f_{XX}-2f_X^2\geq 0,\hspace{.2cm}f_X\geq 0.
\end{equation}
So, close to the  minimum of $f$ it follows that $1-2{\cal{F}}(X-1)\geq0$, which shows that it is sufficient to take ${\cal{F}}\leq\frac{1}{2}$.\\
As for the tensor perturbations, from Eq. (\ref{s2_56}) it turns out that to have well-defined behaiviour we have to impose $z_T^2\geq0$, which is satisfied by Eq. (\ref{report_2}). 
\section{The third-order action}
The primordial non-Gaussianity is negligible for the $R^2$ model, which is in agreement with the observations \cite{{ng-re},{plank9}}. So, it is necessary to study the action (\ref{i-1}) to obtain information about any additional sources for this sector.\\
The leading non-Gaussian signature arises from interactions in the third-order action. From the third-order action, one can obtain the three-point correlation function of the Fourier modes of the curvature perturbation which is related tong-re bi-spectrum, $B(k_1,k_2,k_3)$, as 
\begin{align}
\langle\zeta_{\vk_{1}}\zeta_{\vk_{2}}\zeta_{\vk_3}\rangle=
(2\pi)^3\delta(\vk_1+\vk_2+\vk_3)B_\zeta(k_1,k_2,k_3).
\end{align}
Ref. \cite{plank9} provides constraints on a dimensionless parameter, $f_{NL}$, which is defined as
\begin{equation}
f_{NL}(k_1,k_2,k_3)\equiv\frac{5}{6}\frac{B_\zeta(k_1,k_2,k_3)}{P_\zeta(k_1)P_\zeta(k_2)+2\text{perms.}},
\end{equation}
where perms stands for ``permutations``.\\ 
To find the third-order action, we need $\delta^3R$ in Eq. (\ref{s2_20}), which is given  by Eq. (\ref{apbb3}) in appendix A.\\  
Using the comoving gauge, integrating by parts and the background equations of motion, the third-order action can be obtained as 
\begin{align}
\delta^3S=&\int dt d^3xa^3\\\nn
 &\times\Bigg(\frac{\Mp^2}{2}[1+4\Mp^{-2}Rf]\Bigg[
\frac{2}{a^2}\zeta(\vx,t)(\pa\zeta(\vx,t))^2(\frac{\dot{\th}}{\th^2}
+\frac{H}{\th}-1)\\\nn
&+\frac{\dot{\zeta}(\vx,t)}{\th a^4}(\pa^2\psi_1)^2
-\frac{3}{a^4}\zeta(\vx,t)(\pa^2\psi_1)^2\\\nn
&-\frac{\dot{\zeta}(\vx,t)}{\th a^4}(\pa_i\pa_j\psi_1)(\pa_i\pa_j\psi_1)
-\frac{4}{a^4}\pa^2\psi_1\pa_i\zeta(\vx,t)\pa_i\psi_1\\\nn
&+\frac{3}{a^4}\zeta(\vx,t)(\pa_i\pa_j\psi_1)(\pa_i\pa_j\psi_1)
-2\frac{\dot{\zeta}^3(\vx,t)}{\th^3}\Sigma\\\nn
&+6\frac{\zeta(\vx,t)\dot{\zeta}^2(\vx,t)}{\th^2}\Sigma
\Bigg]\\\nn
&-12X^2R^2\frac{f_X^2}{f}\frac{\zeta(\vx,t)\dot{\zeta}^2(\vx,t)}{\th^2}\\\nn
&+4X^3R^2\Big(2\frac{f_Xf_{XX}}{f}-2\frac{f_X^3}{f^2}-\frac{f_{XXX}}{3}-\frac{f_{XX}}{2X}\Big)\frac{\dot{\zeta}^3(\vx,t)}{\th^3}\Bigg).
\end{align}
At $X=1$, the above action takes the following form
\begin{equation}\label{s5_4}
\delta^3S=\delta^3S|_{R^2}+f{\cal{S}}_f,
\end{equation}
where $\delta^3S|_{R^2}$ is the corresponding action for the $R^2$ model and 
\begin{equation}
{\cal{S}}_f\equiv \frac{f_{XX}}{f}\int dt d^3x 2a^3R^2\l[3\frac{\dot{\zeta}^2(\vx,t){\zeta}(\vx,t)}{\th^2}-2\frac{\dot{\zeta}^3(\vx,t)}{\th^3}\r].
\end{equation}
In Eq. (\ref{s5_4}), $f$ is intentionally considered as the coefficient of ${\cal{S}}_f$.\\
Therefore, in the extended model we have two additional sources for the primordial non-Gaussianity at the minimum.\\
However, during inflation, $\delta^3S|_{R^2}$ is proportional to $f$. So, ${\cal{S}}_f$ is suppressed by $\frac{f_{XX}}{f}$.\\ Using ${\cal{S}}_f$, it is straightforward to calculate the bi-spectrum as Refs. \cite{{ng-re},{see}}. However, even for the $R^2$ model the primordial non-Gaussianity is small enough that we neglect it. So, regarding ${\cal{S}}_f$, we think that such calculations are not necessary for the extend model.

\section{conclusions}
In this work, we proposed an extend $R^2$ model. We showed that the extended $R^2$ model has  similar predictions as the $R^2$ model. Also, the model has new predictions with respect to the $R^2$ model. We have shown that the reason for the similarity of the two models is the large value of the dimensionless parameter of the $R^2$ model. We illustrated these features by comparing the predictions of the models during inflation.\\ 
Also, in the extend model there exist a scalar field which is condensed during inflation. Although additional terms in the cubic action are induced by the scalar field, they do not produce significant non-Gaussiniaty signal.\\
The similarity between the two models will have interesting phenomenological implications. First of all, one can argue that the large value of the parameter is just one of the possible values for the minimums in Fig. 1. So, the extended model can be used to provide a reason for the large value of the parameter.\\
Also, it remains to be seen whether this similarity exists in other eras such as radiation era. If the similarities between the models remain in other eras, then we can regard the $R^2$ model as an example of the extended model. Furthermore, finding any measurable differences between the two models, can be used to test the models.\\
\section{Acknowledgements}
I thank H.~Asgari for useful comments. A. Ghalee is supported by University of Tafresh contract 19-190.
\appendix
\section{The perturbed Ricci scalar}
In the main part of this paper, we have used the following expansion for the Ricci  scalar
\begin{equation}
R=\bar{R}+\delta^1 R+\delta^2R+\delta^3 R+...\hspace{.3cm}.
\end{equation}
Note that we have used the following formula for the Ricci scalar \cite{wald} 
\begin{align}\label{s42_17000}
R=^{(3)}R+K^{ij}K_{ij}-K^2-2\big(n^{\alpha}_{;\beta}n^{\beta}-n^{\alpha}n^{\beta}_{;\beta}\big)_{;\alpha}.
\end{align}
The last term in the above formula is the total derivative term. So, this term has no effect in the Einstein-Hilbert action. However, in our work this term is important. Therefore, compared with other references, reader finds additional terms in the following formulas.\\
The explicit form of the above quantities are
\begin{equation}
\bar{R}=12H^2+6\dot{H},
\end{equation}
\begin{align}\label{apbb1}
\delta^1 R=&-24\alpha_1H^2+24H\dot{\zeta}-8H\pa_iN^i-6H\dot{\alpha_1}
\\\nn
&+6\ddot{\zeta}-2\pa_t\pa_iN^i-12\dot{H}\alpha_1-4a^{-2}\pa^2\zeta
\\\nn
&-2a^{-2}\pa^2\alpha_1,
\end{align}
\begin{align}\label{apbb2}
\delta^2 R=&-24H^2\alpha_2-2a^{-2}\pa^2\alpha_2-6H\dot{\alpha}_2-12\dot{H}
\alpha_2\\\nn
&+\frac{1}{2}\pa_iN^j\pa_iN^j
+8a^{-2}\zeta\pa^2\zeta-2a^{-2}\pa_i\zeta\pa_i\zeta\\\nn
&+4 a^{-2}
\zeta\pa^2\alpha_1
-2a^{-2}\pa_i\zeta\pa_i\alpha_1
+2a^{-2}\alpha_1\pa^2\alpha_1\\\nn
&+36\alpha_1^2H^2-48H\alpha_1\dot{\zeta}
+12\dot{\zeta}^2-6\dot{\alpha_1
}\dot{\zeta}\\\nn
&+18H\alpha_1\dot{\alpha}_1
-12\alpha_1\ddot{\zeta
}+18\alpha_1^2\dot{H}\\\nn
&-12\pa_i\dot{\zeta}N^i+16H\alpha_1\pa_iN^i
-24HN^i\pa_i\zeta\\\nn
&-8\dot{\zeta}\pa_iN^i
+2\dot{\alpha}_1\pa_iN^{i}-6\pa_{i}\zeta\pa_tN^{i}\\\nn
&+4\alpha_1\pa_t\pa_iN^{i}+6HN^i\pa_i\alpha_1+\pa_iN^i\pa_jN^j\\\nn
&+2N^i\pa_i(\pa_jN^j)+\frac{1}{2}\pa_iN^j\pa_jN^i,
\end{align}
\begin{align}\label{apbb3}
\delta^3R=&-8a^{-2}\zeta^2\pa^2\zeta+4a^{-2}\zeta\pa_i\zeta\pa_i\zeta+
72H\alpha_1^2\dot{\zeta}\\\nn
&-24H\alpha_1^2\pa_iN^i-24\alpha_1\dot{\zeta}^2
-\alpha_1\pa_iN^j
\pa_iN^j\\\nn
&-\alpha_1\pa_iN^j\pa_jN^i+48\alpha_1H\pa_i\zeta N^i+16\alpha_1\dot{\zeta}\pa_iN^i\\\nn
&+8\pa_iN^iN^j\pa_j\zeta-24\dot{\zeta}N^i\pa_i\zeta
-2\alpha_1\pa_iN^i\pa_jN^j\\\nn
&-4a^{-2}\zeta^2\pa^2\alpha_1+4\zeta
a^{-2}\pa_i\zeta\pa_i\alpha_1-4\zeta a^{-2}\alpha_1\pa^2\alpha_1\\\nn
&+4a^{-2}\alpha_1\pa_i\alpha_1\pa_i\alpha_1+2a^{-2}\pa_i\zeta\alpha_1
\pa_i\alpha_1+6\dot{\alpha}_1\pa_i\zeta N^i\\\nn
&+18\dot{\zeta}\dot{\alpha}_1
\alpha_1-6\alpha_1\dot{\alpha}_1\pa_iN^i
-36H\dot{\alpha}_1\alpha_1^2\\\nn
&+24\alpha_1\pa_i\dot{\zeta}N^i
+12\alpha_1\pa_i\zeta\pa_tN^i+18\ddot{\zeta}\alpha_1^2\\\nn
&-6\alpha_1^2\pa_t\pa_iN^i-18H\alpha_1\pa_i\alpha_1N^i
+6\dot{\zeta}N^i\pa_i\alpha_1\\\nn
&-2N^j\pa_iN^i\pa_j\alpha_1
+6N^i\pa_i(\pa_j\zeta N^j)\\\nn
&-4\alpha_1N^i\pa_i(\pa_jN^j)
-36H^2\alpha_1^3-24\dot{H}\alpha_1^3\\\nn
&+\text{ terms contain $\alpha_2$ and $\alpha_3$ }\\\nn
&+\pa_i(\text{ third-order terms}),
\end{align}
where $\zeta=\zeta(\vx,t)$.
\bibliographystyle{apsrev4-1}

\end{document}